\documentclass[prl,twocolumn,superscriptaddress,showpacs,a4paper]{revtex4}

\usepackage[dvips]{graphicx}

\begin{document}


\title{Scaling Relations for Logarithmic Corrections}
\date{March 2006}

\author{R.\ Kenna}
\affiliation{Applied Mathematics Research Centre,
Coventry University, Coventry, CV1 5FB, England}

\author{D.A.\ Johnston}
\affiliation{Department of Mathematics, 
School of Mathematical and Computer Sciences,
Heriot-Watt University, Riccarton, Edinburgh EH14 4AS, Scotland}

\author{W.\ Janke}
\affiliation{Institut f\"ur Theoretische Physik, Universit\"at Leipzig,
Augustusplatz 10/11, 04109 Leipzig, Germany}

\begin{abstract}
Multiplicative logarithmic corrections to scaling are frequently encountered in 
the critical behavior of certain statistical-mechanical systems. 
Here, a Lee-Yang
zero approach is used to systematically analyse the exponents of such logarithms
and to propose scaling relations between them.
These proposed relations are then confronted with a variety of results from the literature.
\end{abstract}



\maketitle


Conventional leading scaling behavior at a second-order phase transition 
is described by power laws in the reduced temperature $t$ and field $h$. 
With $h=0$,
the correlation length, specific heat, and susceptibility behave as  
$\xi_\infty(t) \sim |t|^{-\nu}$, $C_\infty(t) \sim |t|^{-\alpha}$, and
$\chi_\infty(t) \sim |t|^{-\gamma}$,  
while the magnetization in the 
broken phase has $m_\infty(t) \sim |t|^{\beta}$. 
Here the subscript indicates the
extent of the system. 
At $t=0$ the magnetization 
scales as $m_\infty(h) \sim h^{1/\delta}$ while the anomalous dimension $\eta$ characterizes
the correlation function at criticality. 
In the 1960's, 
it was shown that these six critical exponents are related via four scaling relations
(see e.g. Ref.~\cite{history} and references therein), which 
are now firmly established and fundamentally important
in the theory of critical phenomena. 
With $d$ representing the dimensionality of the system,
the scaling relations are
\begin{eqnarray}
 \nu d & = & 2-\alpha 
,
\label{Jo}
\\
 2\beta + \gamma & = & 2 - \alpha 
,
\label{Ru}
\\
 \beta (\delta - 1)  & = & \gamma 
,
\label{Gr}
\\
 \nu (2-\eta)  & = & \gamma 
.
\label{Fi}
\end{eqnarray}
In the conventional scaling scenario, 
(\ref{Ru}) and (\ref{Gr}) can, in fact, be deduced from the Widom scaling hypothesis
that the Helmholtz free energy is a homogeneous function \cite{Wi65}.
Widom scaling and the remaining two laws can, in turn, be derived from the Kadanoff 
block-spin construction~\cite{Ka66} and ultimately from Wilson's
renormalization group (RG)~\cite{Wi71}.
The relation (\ref{Jo}) can also be derived from the hyperscaling hypothesis, namely, that the free 
energy behaves near criticality as the inverse correlation volume: 
$f_\infty(t) \sim \xi_\infty^{-d}(t)$.
Twice differentiating this relation recovers (\ref{Jo}).

The  scaling relations, (\ref{Ru}) and (\ref{Gr}), 
were both rederived using an alternative route by Abe \cite{AbeLY} and Suzuki \cite{SuzukiLY}
 exploiting the fact that the even and odd scaling
fields can be linked by Lee-Yang zeros \cite{LY}.
The locus of these zeros in the  magnetic-field plane 
is controlled by the  temperature.
In the $t>0$ (disordered) phase 
this locus terminates at the Yang-Lee edge \cite{LY},
the distance of which from the 
critical point is  denoted by $r_{\rm{YL}}(t)$. 
At a conventional second-order phase transition 
$  r_{\rm{YL}} (t) \sim  t^{\Delta} 
$
for $t>0$,
and the gap exponent $\Delta$ is related to the other 
exponents through \cite{AbeLY,SuzukiLY}
\begin{equation}
\Delta = \frac{\delta \gamma }{\delta - 1}
= \delta \beta = \beta + \gamma .
\label{YLe}
\end{equation}

Logarithmic corrections are characteristic of a number of marginal scenarios 
(see, e.g., Ref.~\cite{We76} and references therein). 
Hyperscaling fails at and above the upper critical dimension $d_c$ 
and, while (\ref{Jo}) holds there, it too fails above $d_c$, where mean-field behavior
(which is independent of $d$) prevails.
At $d_c$ itself, multiplicative logarithmic corrections to scaling
are manifest.
Such corrections are found in marginal $d<d_c$ situations too \cite{We76,BeSh04}. 
The $q$--state Potts
model in $d=2$ dimensions possesses a first-order transition for $q > 4$ and a 
second-order one when $q<4$.
The $q=4$ case is also characterized by
a transition of second order, albeit with multiplicative logarithmic corrections 
to scaling. 
Also in two dimensions, the Ising model with uncorrelated, quenched random-site or random-bond disorder
 offers another example
of such corrections at a demarcation point. 
According to the Harris criterion \cite{Ha74}, when the critical exponent $\alpha$ of the 
specific heat for a pure system is positive, random quenched
disorder is relevant (and exponents may change as disorder is added). If $\alpha$ is
negative in the pure system, the critical behavior is not expected to be altered 
by such disorder. In the marginal case where $\alpha = 0$, no Harris  prediction
can be made, and logarithmic corrections to the pure scaling behavior may ensue.
These are some examples of the rich and disparate variety of systems displaying such phenomena 
and which have been hitherto studied individually. 
Given the ubiquitous role that these logarithms play in 
such marginal cases,
it is reasonable to ask if scaling relations for their exponents exist in analogy to 
(\ref{Jo})--(\ref{YLe}) above.

Here three such relations, together with one for the Yang-Lee edge, 
are derived through the medium of partition function zeros and confronted with the literature.
From the outset we mention that the scaling relations proposed herein do not all
apply to the very special circumstance of the
Ising model in two dimensions and its bond-disordered counterpart. 
These require special treatment beyond the general considerations
presented here. 
For a scaling theory appropriate to the former, see Refs.~\cite{We76,AhFi83} and references therein.
With this in mind, we address the situation with the following scaling behavior:
\begin{eqnarray}
\xi_\infty (t) & \sim & |t|^{-\nu} |\ln{|t|}|^{\hat{\nu}},
\label{xi}
\\
  C_\infty (t) & \sim & |t|^{-\alpha} |\ln{|t|}|^{\hat{\alpha}},
\label{C}
\\
\chi_\infty (t) & \sim & |t|^{-\gamma} |\ln{|t|}|^{\hat{\gamma}},
\label{chi}
\\
  m_\infty (t) & \sim & |t|^{\beta} |\ln{|t|}|^{\hat{\beta}}
\quad {\mbox{for $t<0$}},
\label{mt}
\\
 r_{\rm{YL}}(t) & \sim & t^{\Delta} |\ln{|t|}|^{\hat{\Delta}}
\quad {\mbox{for $t>0$}},
\label{edge}
\end{eqnarray}
while at $t=0$,
\begin{equation}
 m_\infty(h) \sim |h|^{{1}/{\delta}} |\ln{|h|}|^{\hat{\delta}}.
\label{mh}
\end{equation}

In the thermodynamic limit  the free energy 
may be written as
\begin{equation}
 f_\infty (t,h) = 2 {\rm{Re}} \int_{r_{\rm{YL}}(t)}^R{\ln{[h-h(r,t)]}g_\infty(r,t) dr } ,
\label{g1}
\end{equation}
where $R$ is a cutoff, $g_\infty(r,t)$ is the density of zeros, with locus 
$ h(r,t) = r \exp{[i \phi(r,t)]}$. 
If the Lee-Yang circle theo\-rem holds  
the locus is given by 
$\phi=\pi/2$, $R=\pi$ \cite{LY}. 
While the validity of the Lee-Yang circle theorem is
not assumed in what follows (it does not hold for the Potts model, for example),
it is assumed that the small-$t$ critical behavior is dominated by the zeros closest
to the critical point, and that the locus of these zeros can be approximated by $\phi(r,t)  = \phi$,
a constant.

The magnetic susceptibility is the second field derivative of the free energy, 
and, at $h=0$  [substituting $r = xr_{\rm{YL}}(t)$] is
\begin{equation}
 \chi_\infty(t) = -\frac{
                       2\cos{(2\phi)}
                      }{
                       r_{\rm{YL}}(t)
                      } 
\int_1^{{R}/{
                 r_{\rm{YL}}(t)
                 }}{
                            \frac{g_\infty(xr_{\rm{YL}},t)}{x^2}
                         dx }.
\label{g4}
\end{equation}
Expanding (\ref{g4}) about $r_{\rm{YL}}(t)/R=0$ gives
\begin{equation}
 g_\infty(r,t) = 
 \chi_\infty(t) r_{\rm{YL}}(t) 
 \Phi{\left(\frac{r}{r_{\rm{YL}}(t)}\right)},
\label{g5}
\end{equation}
up to additive corrections in $r_{\rm{YL}}(t)/R$ and 
where $\Phi$ is an undetermined function of its argument.
The ratio $r_{\rm{YL}}(t)/R$ is sufficiently 
small near criticality so that these additive corrections may be dropped.
Similar considerations yield for the magnetization
\begin{equation}
 m_\infty(t,h) =  
 \chi_\infty(t)  r_{\rm{YL}}(t)  
\Psi_\phi{\left(\frac{h}{r_{\rm{YL}}(t)}\right)},
\label{g7}
\end{equation}
in which 
\begin{equation}
 \Psi_\phi{\left(\frac{h}{r_{\rm{YL}}(t)}\right)}
=
2 {\rm{Re}}
\int_1^\infty{\frac{\Phi(x) }{h/r_{\rm{YL}}(t)-xe^{i\phi}} dx }.
\label{g6}
\end{equation}
Letting $h \rightarrow 0$ in (\ref{g7}), and comparing to (\ref{mt}),
recovers the scaling relation (\ref{YLe}) and yields 
\begin{equation}
 \hat{\Delta}  =  \hat{\beta} - \hat{\gamma}.
\label{NewLY}
\end{equation}
Furthermore, fixing the argument of the function $\Psi_\phi$ in (\ref{g7}) 
gives $t \sim h^{1/\Delta}|\ln{|h|}|^{-\hat{\Delta}/\Delta}$ from (\ref{edge}),
so that (\ref{g7}) may be written
\begin{equation}
 m_\infty(t,h)   
 \sim h^{1-{\gamma}/{\Delta}}
|\ln{h}|^{\hat{\gamma}+{\gamma \hat{\Delta}}/{\Delta}}
\Psi_\phi{\left(\frac{h}{r_{\rm{YL}}(t)}\right)}.
\label{g77}
\end{equation}
Now taking 
$t \rightarrow 0$ and
comparing with (\ref{mh}) recovers
the known leading behavior for the edge (\ref{YLe}), 
together with 
the correction relation $\hat{\Delta} = \delta (\hat{\delta}-\hat{\gamma})/(\delta - 1)$.
The former
recovers  (\ref{Gr}), while the latter, with (\ref{NewLY}), gives
\begin{equation}
 \hat{\beta} (\delta - 1) =  \delta \hat{\delta} - \hat{\gamma}.
\label{NewW1}
\end{equation}

It is convenient at this point to define the cumulative distribution function of zeros as
\begin{equation}
 G_\infty (r,t) 
  = 
 \int_{r_{\rm{YL}}(t)}^{r}{g_\infty(s,t)ds}
 =  
 \chi_\infty(t) r^2_{\rm{YL}}(t) 
 I{\left(\frac{r}{r_{\rm{YL}}(t)}\right)},
\label{GG}
\end{equation}
in which $I(y) = \int_1^y{\Phi(z)dz}$.
Integrating (\ref{g1}) by parts then gives  the singular part of {\mbox{the free energy 
at $h=0$,}}
$
 f_\infty (t) =  $ $2 \int_{r_{\rm{YL}}(t)}^R{[G_\infty(r,t)/r]} dr
$.
Again substituting $r=xr_{\rm{YL}}(t)$, 
differentiating twice with respect to reduced temperature and comparing the resulting expression 
for the specific heat with (\ref{C})
yields $\alpha = 2 + \gamma - 2\Delta$ and $\hat{\alpha}= \hat{\gamma} + 2 \hat{\Delta}$.
From  (\ref{Gr}) and (\ref{YLe}), 
the first of these is the scaling law (\ref{Ru}).
From (\ref{NewLY}), the second  can be  conveniently expressed as
another relation between the correction exponents, namely
\begin{equation}
 \hat{\alpha} =  2 \hat{\beta} - \hat{\gamma} .
\label{NewW2}
\end{equation}
Using these scaling relations, 
and fixing the ratio $r/r_{\rm{YL}}(t)$ in (\ref{GG})
and then taking the $t\rightarrow 0$ limit, 
gives the critical cumulative distribution function to be 
\begin{equation}
 G_\infty(r,0) \sim r^{{(2-\alpha)}/{\Delta}} |\ln{r}|^{\hat{\alpha}-
 {(2-\alpha)\hat{\Delta}}/{\Delta} }.
\label{GGG}
\end{equation}

Consider now a system of finite extent $L$, and let $h_j(L) = r_j(L)\exp{(i\phi_j)}$ 
be the  $j^{\rm{th}}$ zero there.
The finite-size scaling  (FSS) of first zero is expressible as
\begin{equation}
 \frac{r_1(L)}{r_{\rm{YL}}(t)}
 =
 {\cal{F}} 
 \left(
         \frac{\xi_L(0)}{\xi_\infty(t)}
 \right),
\label{modFSS}
\end{equation}
in which $\xi_L(0)$ is the correlation length of the finite-size system at $t=0$.
On dimensional grounds, we may assume this quantity takes the generic form
\begin{equation}
 \xi_L(0) \sim L (\ln{L})^{\hat{q}},
\label{corrL}
\end{equation}
having allowed for multiplicative logarithmic corrections. 
Recently, additional insights into the origin of FSS were given in Ref.~\cite{JaKe01}.
For a finite system, the cumulative density of zeros is simply the 
fractional number of zeros up to a given point, and we write
\begin{equation}
 G_L(r_j(L)) = \frac{2j-1}{2L^d}.
\label{GL}
\end{equation}
For large enough $L$, and at $t=0$, this must coincide with the expression (\ref{GGG}). 
In particular, it allows the scaling behavior of the lowest 
zero at thermodynamic criticality to be expressed as 
\begin{equation}
 r_1(L) \sim L^{-{d\Delta}/{(2-\alpha)}}
 \left(
  \ln{L}
 \right)^{\hat{\Delta} - {\Delta \hat{\alpha}}/({2-\alpha})}.
\label{Rand}
\end{equation}
Inserting (\ref{xi}), (\ref{edge}), (\ref{corrL}),  and (\ref{Rand}) into (\ref{modFSS})
recovers (\ref{Jo}) and yields a new scaling relation for logarithmic corrections,
namely
\begin{equation}
 \hat{q} = \hat{\nu} + \frac{\nu \hat{\alpha}}{2-\alpha}.
\label{NewK}
\end{equation}
Hyperscaling  corresponds to $\hat{q}=0$.
Relations (\ref{NewW1}) and (\ref{NewW2}) but not (\ref{NewK})  can be derived 
starting with a suitably modified phenomonological Widom  ansatz \cite{Ak01,Ke04}.

To summarize thus far, the
three standard scaling laws (\ref{Jo})--(\ref{Gr}) 
have been recovered and three analogous relations for the logarithmic corrections
(\ref{NewW1}), (\ref{NewW2}), and (\ref{NewK}) presented.
Furthermore, the standard formula (\ref{YLe}) for the edge has been recovered and 
its logarithmic-correction counterpart is given in (\ref{NewLY}).
While the standard scaling laws for the leading critical exponents are well established,
it is now necessary to confront the scaling relations for  corrections with results from the 
literature, and a variety of models with logarithmic corrections 
are examined on a case-by-case basis.


The leading critical exponents for the  $4$-state Potts model in $d=2$ dimensions
were established in Ref.~\cite{leadP} as
$\alpha=2/3$,
$\beta=1/12$,
$\gamma=7/6$, 
$\delta=15$,
and
$\nu=2/3$, and their correction counterparts are \cite{NaSc80CaNa80,SaSo97}
$\hat{\alpha}=-1$,
$\hat{\beta}=-1/8$,
$\hat{\gamma}=3/4$, 
$\hat{\delta}=-1/15$,
and
$\hat{\nu}=1/2$.
FSS of the thermodynamic functions are given in Refs.~\cite{SaSo97,BlEm81},
from which $\hat{q}=0$.
The standard scaling laws, of course, hold and one notes that 
the correction relations (\ref{NewW1}), (\ref{NewW2}), and (\ref{NewK}) hold too,
while (\ref{YLe}) and (\ref{NewLY}) give $\Delta=5/4$ and $\hat{\Delta} = -7/8$ for the edge.
This latter prediction remains to be verified numerically.


The upper critical dimension for  $O(N)$ symmetric $\phi^4_d$ theories
is $d=d_c=4$, where hyperscaling fails
and the leading critical exponents take on their mean-field values,
$\alpha = 0$, $\beta = 1/2$, $\gamma = 1$, $\delta = 3$,
$\nu = 1/2$, $\Delta=3/2$. 
The RG predictions for the corrections are \cite{Ke04,Br82,BLZLuWe89,KeLa91}
$\hat{\alpha}=(4-N)/(N+8)$,
$\hat{\beta} =3/(N+8)$,
$\hat{\gamma}=(N+2)/(N+8)$,
$\hat{\delta}=1/3$,
$\hat{\nu}   =(N+2)/2(N+8)$,
$\hat{\Delta}     =(1-N)/(N+8)$,
$\hat{q}=1/4$, and
all of the correction relations (\ref{NewLY}), (\ref{NewW1}), (\ref{NewW2}), and (\ref{NewK}) 
hold.


The universality class of $O(N)$ spin models can be adjusted  by introducing long-range interactions
decaying as $x^{-(d+\sigma)}$ ($x$ being distance along the lattice),
for which
$d_c=2\sigma$. 
The critical exponents for the $N$-component system were calculated in Ref.~\cite{FiMa72}
and are 
$\alpha=0$, 
$\beta=1/2$,
$\gamma=1$,  
$\delta=3$,
$\nu=1/\sigma$, and obey the leading scaling relations.
The Privman-Fisher form for the free energy  was calculated in Ref.~\cite{LuBl97},
from which the RG predictions 
for the critical exponents could be verified 
and the logarithmic corrections observed. The logarithmic exponents are
$\hat{\alpha}=(4-N)/(N+8)$,
$\hat{\beta} =3/(N+8)$,
$\hat{\gamma}=(N+2)/(N+8)$,
$\hat{\delta}=1/3$,
$\hat{\nu}   =\sigma (N+2)/(N+8)$.
One observes that (\ref{NewW1}) and (\ref{NewW2}) are obeyed, and 
(\ref{NewK}) holds too if $\hat{q}=1/2\sigma$. 
This recovers the known value $\hat{q}=1/4$ for $O(N)\phi^4_4$ theory \cite{Br82} when $\sigma = 2$, 
and leads to
agreement with FSS in the long-range Ising case in two dimensions 
when $\sigma = 1$  \cite{GrHu04}.
Furthermore, (\ref{YLe}) and (\ref{NewLY})
yield  $\Delta=3/2$ and $\hat{\Delta} = (1-N)/(N+8)$ for the edge, which are again identical 
to the $O(N)\phi^4_4$ values. 
The expression (\ref{Rand}) then gives that the first Lee-Yang zero of 
such a system should scale as $r_1(L) \sim L^{-3\sigma/2} (\ln{L})^{-1/4}$.
This predicion for the Lee-Yang zeros of  long-range systems remains to be verified.

Spin glasses, percolation, the  Yang-Lee edge problem and lattice animals 
are all  related to $\phi^3$ field theory.
For each of these  $d_c=6$ except the lattice 
animal problem which has $d_c=8$ \cite{HP}.
Ruiz-Lorenzo gave a compact  description of the scaling  
of the correlation length, susceptibility and specific heat for these  models as \cite{Ru98}
\begin{eqnarray}
\alpha  =  -1, \quad\quad
\gamma = 1, \quad\quad
\nu=\frac{1}{2},\quad\quad \quad
\label{RLMF}
\\
\hat{\alpha}  =  \frac{2(2b-3a)}{4b-a}, \,
\hat{\gamma} = \frac{2a}{4b-a}, \,
\hat{\nu}   =\frac{5a}{6(4b-a)}.
\label{RLcorr}
\end{eqnarray}
The values of $(a,b)$ are 
$(-4m,1-3m)$ for the $m$-component spin glass, 
$(-1,-2)$ for percolation, and 
$(-1,-1)$ for Yang-Lee singularities (which in $d$ dimensions is closely related to the lattice
animal problem in $d+2$ dimensions).
The mean-field values of the critical exponents for spin glasses and percolation  
were calculated in Refs.~\cite{EdAn75HaLu76,Ah80}, respectively, as
$\beta = 1$, $\delta=2$, and, together with (\ref{RLMF}), 
 obey the usual scaling relations (\ref{Jo})--(\ref{Gr}).
The correction exponents (\ref{RLcorr}) satisfy the scaling relations
(\ref{NewW1}) and (\ref{NewW2}) provided that 
$\hat{\beta} = 2 (b-a)/(4b-a)$ and $\hat{\delta}=b/(4b-a)$. In the percolation case
these give $\hat{\beta}=\hat{\delta}=2/7$,
values which are in agreement with explicit calculations \cite{Ah80}.
Also, (\ref{YLe}) and (\ref{NewLY}) now yield 
$\Delta=1$ and $\hat{\Delta}=2(b-2a)/(4b-a)$ for these models, while 
(\ref{NewK}) gives $\hat{q}=1/6$ in each case.
Ruiz-Lorenzo's prediction for this quantity is $\hat{q}=1/3$ \cite{Ru98}
while Ref.~\cite{FoAh04} contains an implicit assumption that $\hat{q}=0$.

The strong universality hypothesis predicts that 
the quenched, disordered  Ising model in $d=2$ dimensions 
has the same leading 
critical exponents as in the pure case with  logarithmic corrections to scaling \cite{DD}.
In particular, Shalaev and later Shankar and Ludwig (SSL) gave
\cite{SSL} 
$\alpha  =  0$, 
$\beta=1/8   $, 
$\gamma = 7/4$,
$\delta=15   $, 
$\nu=1       $, 
$\hat{\alpha} =0            $,
$\hat{\gamma} = 7/8$,
$\hat{\nu}   = 1/2  $,
with the specific heat predicted to be double-logarithmically divergent \cite{DD},
and a more recent RG calculation gave \cite{JuSh96}
$\hat{\beta} =-1/16$ and $\hat{\delta}=0$.

Amongst the SSL values, that for ${\hat{\gamma}}$ of the random-bond version has been the most clearly confirmed \cite{RoAd98}.
While the majority of published opinion  favours  the double-logarithmically divergent specific heat
(see Refs.~\cite{BeSh04,Aade96,TaSh94SeSh98BeCh04} and references therein),
there have been persistent claims in the literature that the specific heat, in fact,
remains finite at criticality in the site-diluted model  \cite{finite,MaKu99}. 
Compatability between the proposed scaling relations and 
the values $\hat{\beta} =-1/16$, $\hat{\gamma} =7/8$, $\hat{\delta}=0$, and $\hat{\nu} =1/2$, 
is established if $\hat{\alpha} = -1$ in this case. 
This value indeed leads to 
a finite specific heat in the random-site version 
and would neatly explain the persistent claims to that effect in the 
literature \cite{finite,MaKu99}
while still being consistent with the strong universality hypothesis. 
These values are also consistent, via (\ref{NewK}), with SSL's ${\hat{\nu}} = 1/2$ 
provided ${\hat{q}} = 0$, a value actually
claimed on the basis of numerical evidence in Ref.~\cite{Aade96} (see also Ref.~\cite{MaKu99}). However,
any value of ${\hat{\nu}}$ and  ${\hat{q}}$, differing by $1/2$, cannot be ruled out on the 
basis of (\ref{NewK}).

Finally, the fact that the logarithmic divergence in the specific heat for the pure Ising model in 
$d=2$ (see Refs.~\cite{We76,AhFi83}) does not directly fit into the scaling scheme proposed here
is to do with special features of that model, which are shared by the random-bond version \cite{SSL,JuSh96}. 
These special features are the vanishing
of the specific-heat exponent $\alpha$ coupled with the property of self-duality
and give rise to an extra logarithmic factor beyond those discussed herein. 
The apparent incompatability in these special cases with the scaling relations 
proposed herein is perhaps a reason why they have gone 
unnoticed as such before.

We wish to thank Bertrand Berche, Juan Ruiz-Lorenzo, and Boris Shalaev
for  correspondences. 
This work was partially supported by the EU RTN-Network ``ENRAGE'': {\em Random Geometry
and Random Matrices: From Quantum Gravity to Econophysics\/} under grant
No.~MRTN-CT-2004-005616.

\vspace{-5mm}
 
%


\begin{thebibliography}{99}

\vspace{-5mm}

\bibitem{history}
M.E.~Fisher, Rev. Mod. Phys. {\bf{70}}, 653 (1998);
D.A.~Lavis and G.M.~Bell, 
{\emph{Statistical Mechanics of Lattice Systems~2\/}}
(Springer, Berlin, 1999).



\bibitem{Wi65}
B.~Widom,~J.~Chem.~Phys.~{\bf{43}},~3892~(1965);~{\bf{43}},~3898 (1965).

\bibitem{Ka66}
L.P.~Kadanoff, Physics (Long Island City, N.Y.) {\bf{2}}, 263 (1966).

\bibitem{Wi71}
K.G.~Wilson,~Phys.~Rev.~B~{\bf{4}},~3174~(1971);~{\bf{4}},~3184~(1971).


\bibitem{AbeLY}
R. Abe, Prog. Theor. Phys. {\bf 38}, 72 (1967). 

\bibitem{SuzukiLY}
M.~Suzuki, Prog. Theor. Phys. {\bf 38}, 289 (1967); {\bf 38}, 1225 (1967).

\bibitem{LY}
C.N.~Yang and T.D.~Lee, Phys. Rev. {\bf{87}},  404 (1952); 
T.D.~Lee and C.N.~Yang, Phys. Rev. {\bf{87}},  410 (1952). 

\bibitem{We76} 
F.J.~Wegner,
in {\emph{Phase Transitions and Critical Phenomena\/}}, 
ed. by C.~Domb and M.S.~Green (Academic Press, London, 1976), Vol.~VI, p.~8.

\bibitem{BeSh04}
B.~Berche and L.N.~Shchur, JETP Lett. {\bf{79}}, 213 (2004).

\bibitem{Ha74}
A.B.~Harris, J. Phys. C {\bf{7}},  1671 (1974). 

\bibitem{AhFi83}
A.~Aharony and M.E.~Fisher, Phys. Rev. B {\bf{27}}, 4394 (1983).

\bibitem{JaKe01}
W.~Janke and R.~Kenna, J. Stat. Phys. {\bf{102}}, 1211 (2001).

\bibitem{Ak01}
N.~Aktekin, J. Stat. Phys. {\bf{104}}, 1397 (2001).

\bibitem{Ke04}
R.~Kenna, Nucl.~Phys.~{\bf{B691}}, 292 (2004).

\bibitem{leadP}
M.P.M.~den~Nijs,~J.~Phys.~A~{\bf{12}},~1857~(1979);
B.~Nienhuis, E.K.~Riedel and M.~Schick, J. Phys. A {\bf{13}},  L189 (1980).

\bibitem{NaSc80CaNa80}
M.~Nauenberg and D.J.~Scalapino,               Phys. Rev. Lett. {\bf{44}}, 837 (1980);
J.L.~Cardy, M.~Nauenberg and D.J.~Scalapino,       Phys. Rev. B  {\bf{22}}, 2560 (1980).

\bibitem{SaSo97}
J. Salas and A.D. Sokal, J. Stat. Phys. {\bf{88}}, 567 (1997).

\bibitem{BlEm81}
J.L.~Black and V.J.~Emery, Phys. Rev. B {\bf{23}}, 429 (1981).

\bibitem{Br82}
E.~Br\'ezin, J. Phys. (Paris) {\bf{43}}, 15 (1982).

\bibitem{BLZLuWe89}
E.~Br\'ezin, J.C.~Le~Guillou and J.~Zinn-Justin, in {\emph{Phase Transitions and Critical Phenomena\/}},
ed. by D.~Domb and M.S.~Green (Academic Press, London, 1976), Vol~VI, p.~127;
M.~L{\"{u}}scher~and~P.~Weisz,~Nucl.~Phys.~{\bf{B318}},~705~(1989).

\bibitem{KeLa91}
R.~Kenna and C.B.~Lang, Phys. Lett. B {\bf{264}}, 396 (1991);
                        Nucl.~Phys.~{\bf{B393}}, 461 (1993); 
                                    {\bf{B411}}, 340 (1994). 

\bibitem{FiMa72} 
M.E.~Fisher, S.-K.~Ma and B.G.~Nickel, Phys. Rev. Lett. {\bf{29}}, 917 (1972).

\bibitem{LuBl97}
E.$\,$Luijten~and~H.W.J.$\,$Bl{\"{o}}te,$\,$Phys.$\,$Rev.$\,$B$\,${\bf{56}},$\,$8945$\,$(1997).

\bibitem{GrHu04}
D.~Gr{\"{u}}neberg and A. Hucht, Phys. Rev. E {\bf{69}}, 036104 (2004).

\bibitem{HP}
H.-P.~Hsu, W.~Nadler and P.~Grassberger, J.~Phys.~A {\bf{38}}, 775 (2005).


\bibitem{Ru98}
J.J.~Ruiz-Lorenzo, J.~Phys.~A {\bf{31}}, 8773 (1998).

\bibitem{EdAn75HaLu76} 
S.F.~Edwards and P.W.~Anderson,                  J. Phys. F {\bf{5}}, 965 (1975);
 A.B.~Harris, T.C.~Lubensky and J.-H.~Chen, Phys. Rev. Lett. {\bf{36}}, 415 (1976).

\bibitem{Ah80}
A.B.~Harris, J.C.~Lubensky, W.K.~Holcomb and C.~Dasgupta, Phys. Rev. Lett. {\bf{35}}, 327 (1975);
I.W.~Essam,  D.S.~Gaunt and A.J.~Guttmann,                J. Phys. A {\bf{11}}, 1983 (1978);
A.~Aharony,                                               Phys. Rev. B {\bf{22}}, 400 (1980).

\bibitem{FoAh04}
S.~Fortunato, A.~Aharony, A.~Coniglio and D.~Stauffer, Phys. Rev. E {\bf{70}}, 056116 (2004).


\bibitem{DD}
Vik.~S.~Dotsenko and Vl.~S.~Dotsenko, JETP Lett. {\bf{33}}, 37 (1981);
                                     Adv. Phys. {\bf{32}}, 129 (1983).


\bibitem{SSL}
B.N.~Shalaev,  Sov. Phys. Solid State {\bf{26}},  1811 (1984);
               Phys. Rep.             {\bf{237}},  129 (1994);
R.~Shankar,    Phys.~Rev.~Lett.       {\bf{58}},  2466 (1987); 
                                      {\bf{61}},  2390 (1988);
A.W.W.~Ludwig,~Phys. Rev.~Lett.~{\bf{61}},~2388~(1988);~Nucl.~Phys.~{\bf{B330}},~639~(1990).

\bibitem{JuSh96}
G.~Jug and B.N.~Shalaev, Phys. Rev. B {\bf{54}},  3442 (1996).

\bibitem{RoAd98}
A.~Roder, J.~Adler and W.~Janke, Phys. Rev. Lett. {\bf{80}}, 4697 (1998);
                                 Physica (Amsterdam)       {\bf{265A}},  28 (1999). 

\bibitem{Aade96}
F.D.A.~Aar\~{a}o Reis, S.L.A.~de~Queiroz and R.R.~dos~Santos, Phys. Rev. B {\bf{54}}, R9616 (1996);
                                                       {\bf{56}},  6013 (1997).

\bibitem{TaSh94SeSh98BeCh04}
A.L.~Talapov and L.N.~Shchur, J. Phys. Condens. Matter {\bf{6}}, 8295 (1994);
V.N.~Plechko, Phys. Lett.~A {\bf{239}}, 289 (1998);
W.~Selke, L.N.~Shchur and O.A.~Vasilyev,      Physica (Amsterdam) {\bf{259A}}, 388 (1998);
B.~Berche and C.~Chatelain,
in {\emph{Order, Disorder and Criticality\/}}, 
ed. by Yu.~Holovatch (World Scientific, Singapore, 2004), p.~146.

\bibitem{finite} 
K.~Ziegler, J.~Phys. A {\bf{21}}, L661 (1988); 
Nucl. Phys. {\bf{B344}}, 499 (1990);
Europhys. Lett. {\bf{14}}, 415 (1991); J.-K.~Kim and
 A.~Patrascioiu, Phys. Rev. Lett. {\bf{72}}, 2785 (1994);
                              Phys. Rev. B     {\bf{49}}, 15764 (1994);
J.-K.~Kim, cond-mat/9502053;
           Phys. Rev. B                         {\bf{61}},  1246 (2000);
R.~K{\"{u}}hn, Phys. Rev. Lett. {\bf{73}}, 2268 (1994); 
A.C.D.~van~Enter, C.~K{\"{u}}lske and C.~Maes, {\emph{ibid\/}}.  {\bf{84}}, 6134 (2000);
 R.~K{\"{u}}hn and G.~Mazzeo,              {\emph{ibid\/}}. {\bf{84}}, 6135 (2000).


\bibitem{MaKu99} 
G.~Mazzeo and R. K{\"{u}}hn, Phys. Rev. E {\bf{60}}, 3823 (1999).



\end{thebibliography}
\end{document}